\documentclass[12pt,a4paper,notitlepage]{article}  
\usepackage{epsfig}

\def\lsim{\mathrel{\rlap{\lower4pt\hbox{\hskip1pt$\sim$}}
    \raise1pt\hbox{$<$}}}         %less than or approx. symbol
\def\gsim{\mathrel{\rlap{\lower4pt\hbox{\hskip1pt$\sim$}}
    \raise1pt\hbox{$>$}}}         %greater than or approx. symbol
\newcommand{\neu}{\ensuremath{\tilde{\chi}}}

\begin{document}
\def\esim{\mathrel{\rlap{\raise2pt\hbox{$\sim$}}
    \lower1pt\hbox{$-$}}}         %equal to or approx. symbol
\newcommand{\bsg}{${\rm b}\to {\rm s}\gamma$}
\newcommand{\beq}{\begin{equation}}
\newcommand{\eeq}{\end{equation}}

%%%%%%%%%%%%%%%%%%%%%%%%%%%%%%%%%%%%%%%%%%%%%%%%%%%%%%%%%%%%%%%%%%%%%
%                      TITLE PAGE + ABSTRACT
%%%%%%%%%%%%%%%%%%%%%%%%%%%%%%%%%%%%%%%%%%%%%%%%%%%%%%%%%%%%%%%%%%%%%

\rightline{UUITP-13/96}
\rightline{OUTP-96-29P}
\rightline{hep-ph/9607237}

\vspace{1cm}

\begin{center}

{\Large \bfseries Indirect Neutralino Detection Rates \\[1ex]  in Neutrino
  Telescopes}\\

\vspace{1cm}

{\large Lars Bergstr{\"o}m\footnote{Permanent address: Department of Physics,
Stockholm University, Box 6730, SE-113 85 Stockholm, Sweden; E-mail
address: lbe@physto.se}
\qquad
Joakim Edsj{\"o}\footnote{E-mail address: edsjo@teorfys.uu.se}\\
{\small \em Department of Theoretical Physics, Uppsala University,}\\ 
{\small \em Box 803, SE-751 08 Uppsala, Sweden}\\[2ex]

Paolo Gondolo\footnote{E-mail address: p.gondolo1@physics.oxford.ac.uk} \\
{\small \em Department of Physics, University of Oxford,} \\
{\small \em 1 Keble Road, Oxford, OX1 3NP, United Kingdom}}\\

\end{center}

\vspace{0.5cm}

\begin{abstract}  
\smallskip
Neutralinos annihilating in the center of the Sun or the Earth may
give rise to a detectable signal of neutrinos. We derive the indirect
detection rates for neutrino telescopes in the minimal supersymmetric
extension of the standard model. We show that even after imposing
all phenomenological and experimental constraints that make the theories
viable, regions of parameter space exist which can already be probed
by existing neutrino telescopes. We compare with the discovery potential
of supersymmetry at LEP2 as well as direct detections and point out the
complementarity of the methods.
\end{abstract}

%%%%%%%%%%%%%%%%%%%%%%%%%%%%%%%%%%%%%%%%%%%%%%%%%%%%%%%%%%%%%%%%%%%%%
%                     BEGINNING OF TEXT
%%%%%%%%%%%%%%%%%%%%%%%%%%%%%%%%%%%%%%%%%%%%%%%%%%%%%%%%%%%%%%%%%%%%%

%Phys. Lett. B formatting
%\setlength{\baselineskip}{2.0\baselineskip}

\section{Introduction} \label{sec:intro}

Supersymmetric neutralinos with masses in the GeV--TeV range are among
the leading non-baryonic candidates for the dark matter in our
galactic halo. One of the most promising methods for the discovery of
neutralinos  in
the halo is via observation of energetic neutrinos from their annihilation
in the Sun \cite{neusun} and/or the Earth \cite{neuea}.  Through elastic
scattering with the atomic nuclei in the Sun or the Earth, a
neutralino from the halo
can lose enough energy to remain gravitationally trapped \cite{trap}. Trapped
neutralinos sink to the core of the Sun or the Earth where they
annihilate into
ordinary particles: leptons, quarks, gluons and -- depending on the masses --
Higgs and gauge bosons. Because of absorption in the solar or terrestrial
medium, only neutrinos are capable of escaping to the
surface. Neutralinos do not
annihilate into neutrinos directly \cite{Goldberg}, but energetic
neutrinos may be produced via hadronization and/or
decay  of the direct
annihilation products. These energetic neutrinos may be discovered by
terrestrial neutrino detectors.

In this letter, we consider \v{C}erenkov neutrino telescopes. They consist of
large underground arrays of photo-multipliers to detect the \v{C}erenkov light
emitted by muons generated in charged-current interactions of neutrinos with
the medium surrounding the detector. Underground \v{C}erenkov detectors,
originally built to search for proton decay, have already started to explore
(and constrain)  neutralinos as dark matter candidates
\cite{WIMPbounds,Baksan}.
A new generation of much larger neutrino telescopes, utilizing large volumes
of natural water or ice (for a review, see \cite{barwick94}) is currently
under construction, which will increase the sensitivity for high-energy 
neutrino point sources by an order of magnitude or more. 

The prediction of muon rates is quite involved: we compute
neutralino capture rates in the Sun and the Earth, fragmentation
functions in basic annihilation processes, propagation through the
solar or terrestrial medium, charged current cross sections and muon
propagation in the rock, ice or water surrounding the detector. In
addition, there may be scattering of the \v{C}erenkov photons
generated by the muons, due to impurities in the medium.

\section{Definition of the Supersymmetric Model} \label{sec:susy}

The minimal $N=1$ supersymmetric extension of the standard model is
defined by the following superpotential and soft supersymmetry-breaking
potential (for notation and details, see Ref.~\cite{bsg})
\begin{equation}
  W = \epsilon_{ij} \left(
  - {\bf \hat{e}}_{R}^{*} {\bf Y}_E {\bf \hat{l}}^i_{L} {\hat H}^j_1 
  - {\bf \hat{d}}_{R}^{*} {\bf Y}_D {\bf \hat{q}}^i_{L} {\hat H}^j_1 
  + {\bf \hat{u}}_{R}^{*} {\bf Y}_U {\bf \hat{q}}^i_{L} {\hat H}^j_2 
  - \mu {\hat H}^i_1 {\hat H}^j_2 
  \right),
  \label{superpotential}
\end{equation}
\begin{eqnarray}
  \label{Vsoft}
  V_{{\rm soft}} & = & 
  \epsilon_{ij} \left(
  - {\bf \tilde{e}}_{R}^{*} {\bf A}_E {\bf Y}_E {\bf \tilde{l}}^i_{L} H^j_1 
  - {\bf \tilde{d}}_{R}^{*} {\bf A}_D {\bf Y}_D {\bf \tilde{q}}^i_{L} H^j_1 
  + {\bf \tilde{u}}_{R}^{*} {\bf A}_U {\bf Y}_U {\bf \tilde{q}}^i_{L} H^j_2
  \right. \nonumber \\ && \phantom{\epsilon_{ij} \Bigl(} \left.
  - B \mu H^i_1 H^j_2 + {\rm h.c.} 
  \right) \nonumber \\ &&
  + H^{i*}_1 m_1^2 H^i_1 + H^{i*}_2 m_2^2 H^i_2
  \nonumber \\ && +
  {\bf \tilde{q}}_{L}^{i*} {\bf M}_{Q}^{2} {\bf \tilde{q}}^i_{L} + 
  {\bf \tilde{l}}_{L}^{i*} {\bf M}_{L}^{2} {\bf \tilde{l}}^i_{L} + 
  {\bf \tilde{u}}_{R}^{*} {\bf M}_{U}^{2} {\bf \tilde{u}}_{R} + 
  {\bf \tilde{d}}_{R}^{*} {\bf M}_{D}^{2} {\bf \tilde{d}}_{R} + 
  {\bf \tilde{e}}_{R}^{*} {\bf M}_{E}^{2} {\bf \tilde{e}}_{R} 
  \nonumber \\ && +
  {1\over2} M_1 \tilde{B} \tilde{B} +
  {1\over2} M_2 \left( \tilde{W}^3 \tilde{W}^3 +
        2 \tilde{W}^+ \tilde{W}^- \right) +
  {1\over2} M_3 \tilde{g} \tilde{g} . 
\end{eqnarray}
Here $i$ and $j$ are SU(2) indices ($\epsilon_{12} = +1$), ${\bf
Y}$'s, ${\bf A}$'s and ${\bf M}$'s are $3\times3$ matrices in
generation space, and the other boldface letters are vectors in
generation space. 
  
Electroweak symmetry breaking is caused by both $H^1_1$ and $H^2_2$
acquiring vacuum expectation values,
\begin{equation}
  \langle H^1_1\rangle = v_1 , \qquad \langle H^2_2\rangle = v_2,
\end{equation}
with $g^2(v_1^2+v_2^2) = 2 m_W^2$, with the further assumption that
vacuum expectation values of all other scalar fields (in particular,
squark and sleptons) vanish. This avoids color and/or charge breaking
vacua.

To reduce the number of free parameters, we make simplifying
unification assumptions for the gaugino mass parameters
\begin{eqnarray}
  M_1 & = & {5\over 3}\tan^2\theta_wM_2,\\
  M_2 & = & { \alpha_{em} \over \sin^2\theta_w \alpha_s } M_3,
\end{eqnarray}
and a simple ansatz for the soft supersymmetry-breaking parameters in
the sfermion sector:
\begin{eqnarray}
  \mbox{\bf M}_Q & = & \mbox{\bf M}_U = \mbox{\bf M}_D = \mbox{\bf M}_E =
   \mbox{\bf M}_L = m_0 \mbox{\bf 1} , \\
  {\bf A}_U & = & \mbox{\rm diag}(0,0,A_t) , \\ 
  {\bf A}_D & = & \mbox{\rm diag}(0,0,A_b) , \\
  {\bf A}_E & = & 0 .
\end{eqnarray}

We choose as independent parameters the mass $m_A$ of the CP-odd Higgs
boson, the ratio of Higgs vevs $\tan\beta=v_2/v_1$, the gaugino mass
parameter $M_2$, the higgs(ino) mass parameter $\mu$, and the
quantities $m_0$, $A_t$ and $A_b$ above.

We include one-loop radiative corrections and two-loop leading-log
contributions to the Higgs mass matrices
using the effective potential approach described in \cite{Carena}.

The neutralinos $ \tilde{\chi}^0_i$ are linear combinations of the
neutral gauginos ${\tilde B}$, ${\tilde W_3}$ and of the neutral
higgsinos ${\tilde H_1^0}$, ${\tilde H_2^0}$.  In this basis, their
mass matrix
\begin{eqnarray}
  {\cal M}_{\tilde \chi^0_{1,2,3,4}} = 
  \left( \matrix{
  {M_1} & 0 & -{g'v_1\over\sqrt{2}} & +{g'v_2\over\sqrt{2}} \cr
  0 & {M_2} & +{gv_1\over\sqrt{2}} & -{gv_2\over\sqrt{2}} \cr
  -{g'v_1\over\sqrt{2}} & +{gv_1\over\sqrt{2}} & 0 & -\mu \cr
  +{g'v_2\over\sqrt{2}} & -{gv_2\over\sqrt{2}} & -\mu & 0 \cr
  } \right)
\end{eqnarray}
can be diagonalized analytically to give four neutral Majorana states,
\begin{equation}
  \tilde{\chi}^0_i = 
  N_{i1} \tilde{B} + N_{i2} \tilde{W}^3 + 
  N_{i3} \tilde{H}^0_1 + N_{i4} \tilde{H}^0_2 ,
\end{equation}
the lightest of which, to be called $\chi$, is then the candidate for
the particle making up (at least some of) the dark matter in the universe.

We consider only models that satisfy all accelerator constraints on
supersymmetric particles and couplings; in particular, the measurement
of the \bsg\ process at the Cornell accelerator \cite{CLEO}, which
provides important bounds. However, to show the impact of present dark
matter searches, we plot also models which are excluded by these
non-accelerator searches.

%%%%%%%%%%%%%%%%%%%%%%%%%%%%%%%%%%%%%%%%%%%%%%%%%%%%%%%%%%%%%%%%%%%%%%%
\section{Relic Density and Capture Rates} \label{sec:reldens}

For each model allowed by the accelerator constraints we calculate the
relic density of neutralinos $\Omega_\chi h^2$. We use the formalism
in Ref.~\cite{GondoloGelmini} to carefully treat resonant
annihilations and threshold effects, keeping finite widths of unstable
particles, including all two-body annihilation channels of
neutralinos.  The annihilation cross sections were derived using a
novel helicity projection technique \cite{paolo}, and were checked
against published results for several of the subprocesses. 

In this letter, we keep only models in which the neutralino density
does not overclose the Universe and in which neutralinos can make up
the totality of the galactic dark matter. Namely we require
$\Omega_{\rm gal.DM} h^2 < \Omega_\chi h^2 < 1$, where (somewhat
arbitrarily) we choose $\Omega_{\rm gal.DM} h^2 = 0.025$. We have
adopted a local dark matter density of 0.3 GeV/cm$^3$.

The capture rate in the Earth is dominated by scalar interactions, and
presents kinematic enhancements whenever the mass of the neutralino
almost matches one of the heavy elements in the Earth.  For the Sun,
both axial interactions with hydrogen and scalar interactions with
heavier elements are important.  For both the Sun and the Earth we use
the convenient approximations available in \cite{JKG}.

%%%%%%%%%%%%%%%%%%%%%%%%%%%%%%%%%%%%%%%%%%%%%%%%%%%%%%%%%%%%%%%%%%%%%

\section{Muon fluxes from neutralino annihilations} \label{sec:achann}

Neutralinos in the core of the Sun and/or Earth can annihilate to a
fermion-antifermion pair, to gauge bosons, Higgs bosons and gluons
($\chi\chi \to \ell^+\ell^-$, $q\bar{q}$, $gg$, $q\bar{q}g$, $W^+W^-$,
$Z^0Z^0$, $Z^0H^0$, $W^{\pm}H^{\mp}$, $H^0H^0$). These annihilation
products will hadronize and/or decay, eventually producing high energy
muon neutrinos.  Since the rate of muons in a neutrino telescope is
approximately proportional to the neutrino energy squared (since both
the cross section and the muon range are approximately proportional to
the energy), the annihilation channels with the hardest neutrino
spectra will be the most important, i.e.\ $W^+W^-$, $Z^0Z^0$, $t \bar{t}$,
etc. In our calculation of the neutrino fluxes we have however
included all annihilation channels (except gluons since they give very
soft neutrino spectra).

With Monte Carlo simulations we have considered the whole chain of 
processes from the annihilation products in the core of the Sun or the
Earth to detectable muons at the surface of the Earth \cite{Edpre}. 
We have performed a full Monte Carlo simulation of the
hadronization and decay of the annihilation products using {\sc Jetset} 7.4
\cite{Jetset}, of the neutrino interactions on their way out of the Sun and of
the charged-current neutrino interactions near the detector using {\sc Pythia}
5.7 \cite{Jetset}, and finally of the multiple Coulomb scattering of the muon
on its way to the detector using distributions from Ref.~\cite{PDG}.

With respect to calculations using Ref.~\cite{RS} (e.g.\
Ref.~\cite{neuprod}),this 
Monte Carlo treatment of the neutrino propagation through the Sun
bypasses simplifying assumptions previously made, namely neutral
currents are no more assumed to be much weaker than charged currents
and energy loss is no more considered continuous. For details on this
treatment, see Ref.~\cite{Edpre,Angdist}.

The muon flux at a detector has been simulated for a set of neutralino
masses ($m_{\neu}$ = 10, 25, 50, 80.3, 91.2, 100, 150, 176, 200, 250,
350, 500, 750, 1000, 1500, 2000, 3000 and 5000 GeV) and 
annihilation channels ($c\bar{c}$, $b\bar{b}$, $t\bar{t}$, $\tau^+\tau^-$,
$W^+W^-$ and $Z^0Z^0$). For each mass and channel, $2.5 \times 10^5$
annihilations have been simulated. For masses other than those
simulated, an interpolation is performed and the muon flux from
channels other
than those listed above are easily calculated since all other annihilation
products decay to these particles (lighter quarks, electrons and muons
do not contribute significantly to the neutrino flux). For the Higgs
bosons, which
decay in flight, an integration over the angle of the decay products
with respect to the direction of the Higgs boson is performed. Given
the branching ratios for different annihilation channels it is then
straightforward to obtain the muon flux above any given energy threshold
and within any angular region around the Sun or the center of the Earth.

%%%%%%%%%%%%%%%%%%%%%%%%%%%%%%%%%%%%%%%%%%%%%%%%%%%%%%%%%%%%%%%%%%%%%
\section{Indirect detection rates}
\label{sec:indrates}

To illustrate the potential of neutrino telescopes for discovery of
dark matter through neutrinos from the Earth or the Sun, we present
the results of our full calculation.  We show together results
obtained with one `normal' scan in the parameter space letting $\mu$,
$M_2$, $\tan \beta$, $m_A$, $m_0$, $A_b$ and $A_t$ vary at random
between generous bounds and one `special' scan where we have been more
restrictive on the $A$ mass,
\begin{displaymath}
\left\{ \begin{array}{l}
  \mu \in [-5000,5000] \mbox{ GeV} \\
  M_2 \in [-5000,5000] \mbox{ GeV} \\
  \tan \beta \in [1.2,50] \\
  m_{A} \in [0,1000] \mbox{ GeV}\\
  m_0 \in [100,3000] \mbox{ GeV}\\
  A_b \in [-3,3] m_0 \\
  A_t \in [-3,3] m_0 
\end{array}\right. \quad \mbox{`normal'}
\qquad
\left\{ \begin{array}{l}
  \mu \in [-5000,5000] \mbox{ GeV} \\
  M_2 \in [-5000,5000] \mbox{ GeV} \\
  \tan \beta \in [1.2,50] \\
  m_{A} \in [0,150] \mbox{ GeV}\\
  m_0 \in [100,3000] \mbox{ GeV}\\
  A_b \in [-3,3] m_0 \\
  A_t \in [-3,3] m_0 
\end{array}\right. \quad \mbox{`special'}
\end{displaymath}
We recall that the density of points in the figures reflects our
choices for scanning the parameter space, and is therefore subjective
(for a discussion on this see Ref.~\cite{bsg}).

\begin{figure}[h]
  \epsfig{file=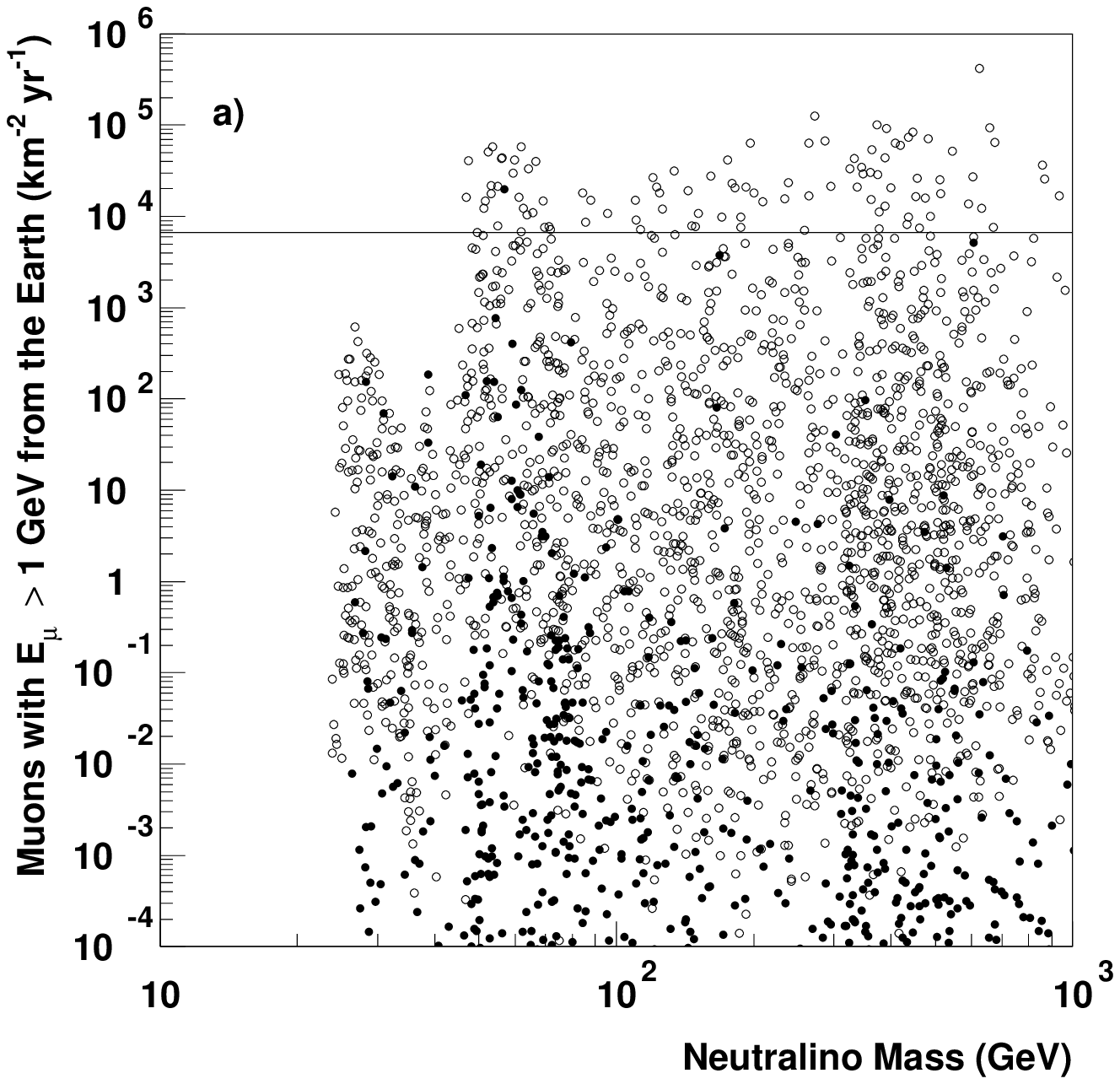,width=\textwidth}
  \caption{The indirect detection rates from
  neutralino annihilations in a) the Earth and b) the Sun versus the
  neutralino mass. The
  horizontal line is the Baksan limit \protect\cite{Baksan}. Filled
  points are from a `normal' scan and open points are from a `special'
  scan with $m_{A}<150$ GeV.}
  \label{fig:rvsmx}
\end{figure}

 In Fig.~\ref{fig:rvsmx} we show our predictions for the indirect
detection rates as a function of neutralino mass. The horizontal lines
are the best present limits for indirect searches and come from the
Baksan detector \cite{Baksan}. The limits are $\Phi_\mu^{Earth} < 2.1
\times 10^{-14}$ cm$^{-2}$ s$^{-1}$ and $\Phi_\mu^{Sun} < 3.5 \times
10^{-14}$ cm$^{-2}$ s$^{-1}$ at 90\% confidence level and integrated
over a half-angle aperture of 30$^\circ$ with a muon energy threshold
of 1 GeV.  Quite a few models with high neutrino rates are already
ruled out by the Baksan data. These models have large $\tan\beta \gsim
30$, low $H_2$ mass,
 and large mixing for stop squarks. In Fig.~\ref{fig:rvsmx} it can
also be seen that a neutrino telescope of an area around 1 km$^2$,
which is a size currently being discussed, would have a large
discovery potential for supersymmetric dark matter. 
We remind that, after subtraction of the
atmospheric neutrino fluxes (by means of an on-source off-source
difference, {\it e.g.}), the remaining background is due to
 high-energy neutrinos produced by cosmic ray collisions in the
solar atmosphere, and is at the level of 15 muons ($>1$ GeV)/km$^2$/yr
\cite{SeckelStanevGaisser}.

\begin{figure}[h]
  \epsfig{file=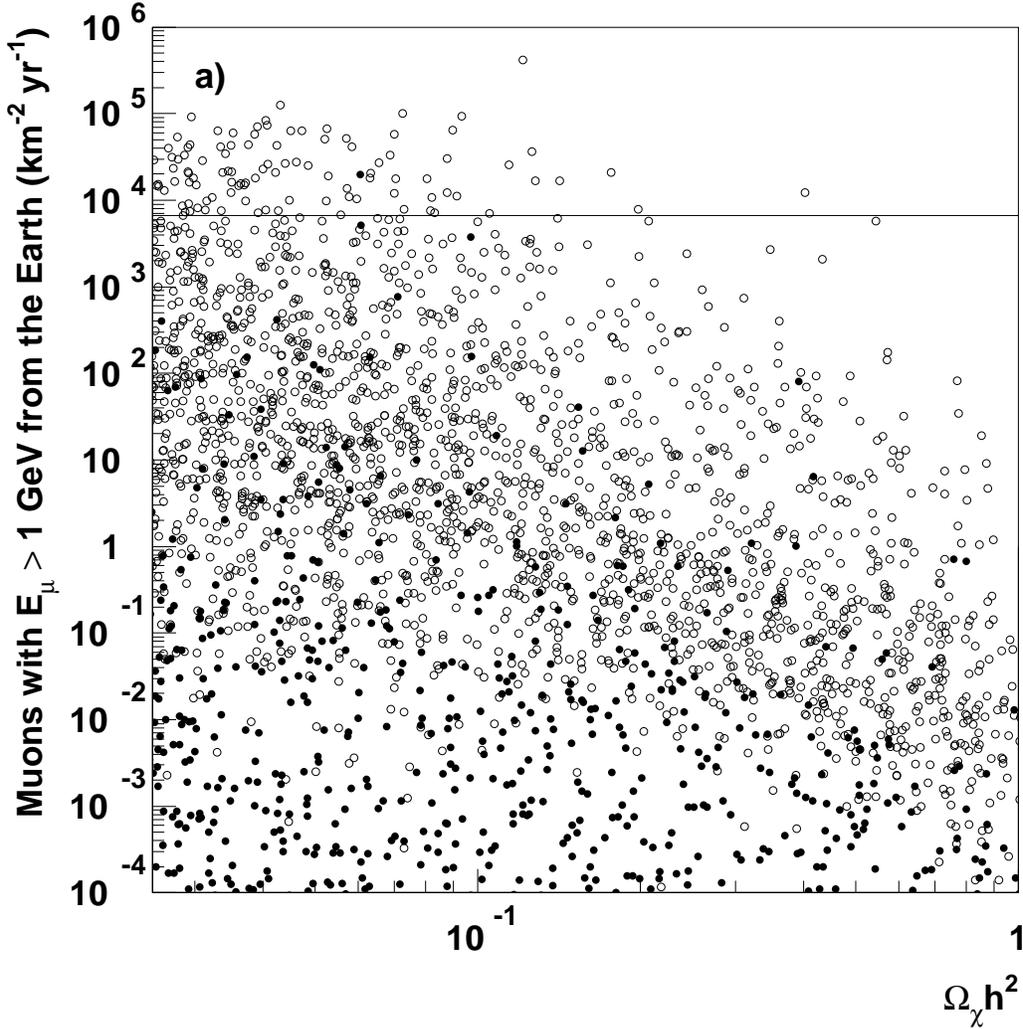,width=\textwidth}
  \caption{The indirect detection rates from neutralino annihilations
    in a) the Earth and b) the Sun versus $\Omega_{\chi} h^2$\@. The
  horizontal line is the Baksan limit \protect\cite{Baksan}. Filled
  points are from a `normal' scan and open points are from a `special'
  scan with $m_{A}<150$ GeV.} 
  \label{fig:rvsoh2}
\end{figure}

In Fig.~\ref{fig:rvsoh2} we show the muon rates versus $\Omega_\chi
h^2$. The general trend is that large $\Omega_\chi h^2$ corresponds to
lower rates, as would be expected from crossing symmetry between
annihilation and scattering cross sections. Note, however, the large
spread of the predicted rates for a given value of $\Omega_\chi
h^2$\@. 

\begin{figure}[h]
  \epsfig{file=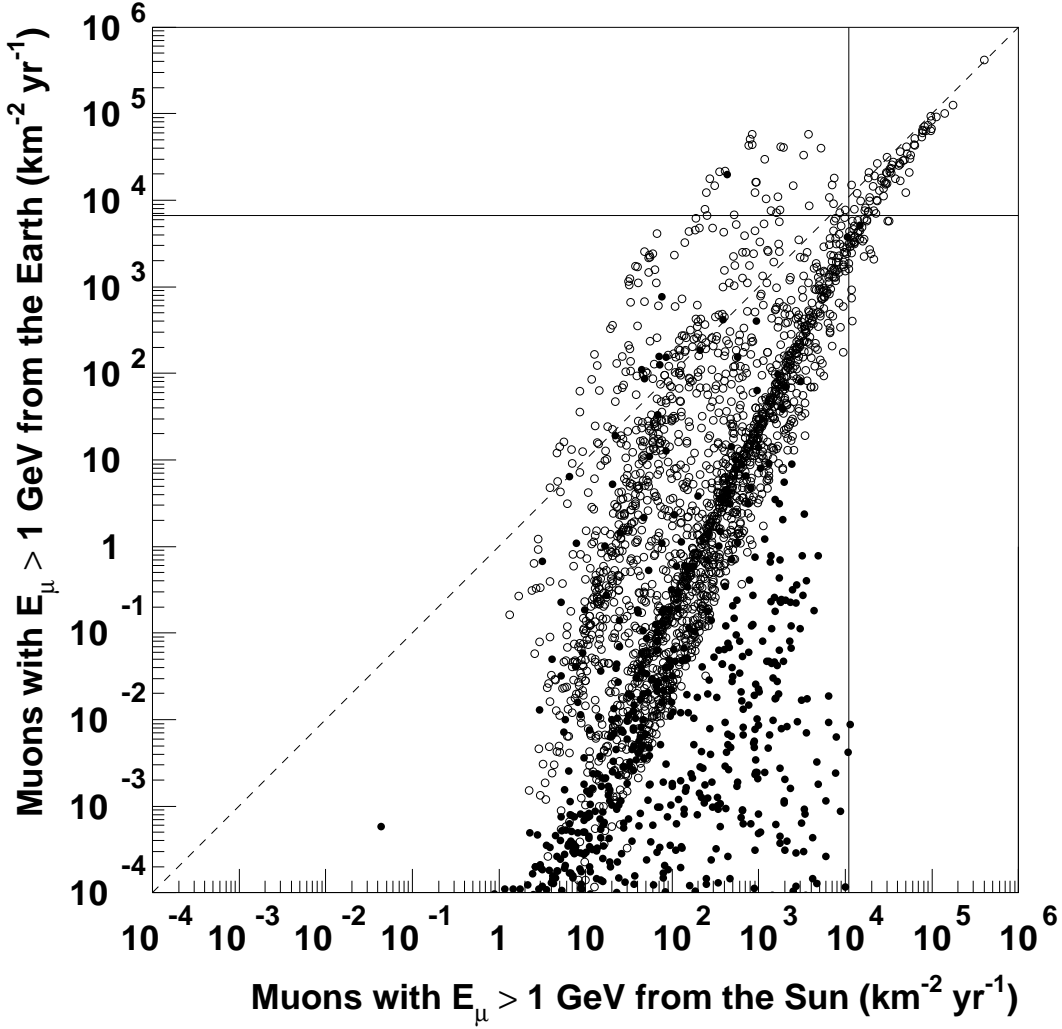,width=\textwidth}
  \caption{The indirect detection rates in the Earth versus the
  indirect detection rates in the Sun. The horizontal and vertical
  lines are the Baksan limits \protect\cite{Baksan}. The dashed line,
  indicating equal rates, is shown just for convenience. Filled
  points are from a `normal' scan and open points are from a `special'
  scan with $m_{A}<150$ GeV.}
  \label{fig:reavsrsu}
\end{figure}

In Fig.~\ref{fig:reavsrsu} a comparison is made between the predicted
rates from the Earth and from the Sun for the same set of models. In
the region yet to explore, the rate from the Sun is generally larger
than the rate from the Earth.

\begin{figure}[h]
  \epsfig{file=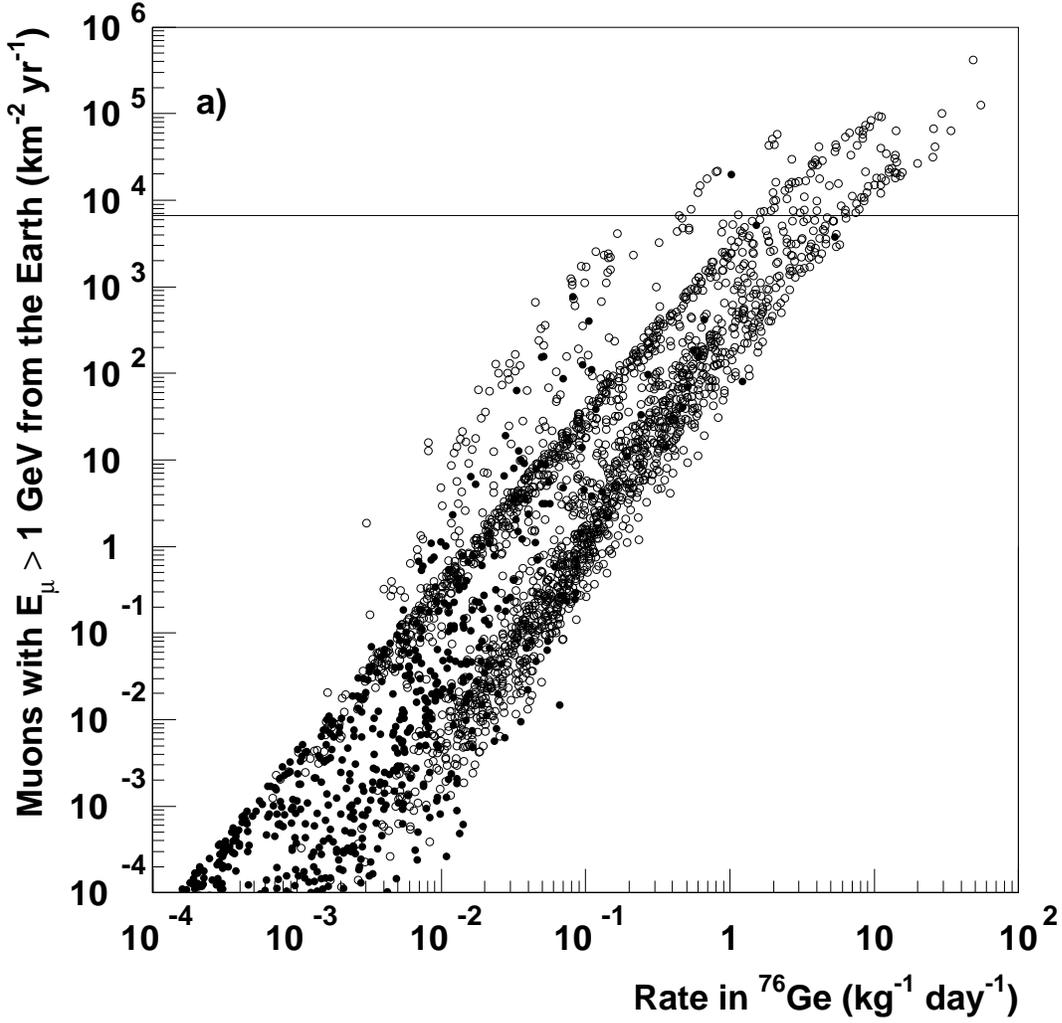,width=\textwidth}
  \caption{The indirect detection rates from neutralino annihilations in 
  a) the Earth and b) the Sun versus the direct detection rate in $^{76}$Ge. 
  The horizontal line is the Baksan limit \protect\cite{Baksan}. Filled
  points are from a `normal' scan and open points are from a `special'
  scan with $m_{A}<150$ GeV.} 
  \label{fig:rvsrge}
\end{figure}

In Fig.~\ref{fig:rvsrge} the indirect detection rate is compared to
the direct detection rate in $^{76}$Ge. As can be seen, there is a
correlation between the two, although for the Sun it is not as strong
as for the Earth, where a high capture rate is due to a large scalar
cross section, which also means a high rate in Germanium\footnote{Recently, 
a paper appeared (V.A.~Bednyakov et al., hep-ph/9606261),
where no high Germanium rates were found in a variant of the special scan 
employed here. This seems to be due to their use of a more
restricted model, imposing unification conditions of the scalar
mass parameters at the GUT scale, which we do not.}. 
Without
forgetting the huge spread, we see that
for a given factor of improvement in sensitivity, indirect
detection from the Sun generally scores better than direct detection,
which in turn generally scores better than indirect detection from the
Earth.

Note that the muon rates in real experiments may be significantly
lower (by as much as an order of magnitude for neutralinos in the lower
mass range) due to the need 
to impose a higher
energy threshold  for the signal than assumed here. We have taken 1 GeV
for neutrino telescopes which is true for a small scale detector like
Baksan; for a kilometer-scale array it is more likely to be tens of GeV\@. 
Likewise, the Germanium rate given is the integrated rate from zero recoil
energy
to the kinematical limit. Present-day detectors typically only sample a
small range of recoil energies. 

\begin{figure}[h]
  \epsfig{file=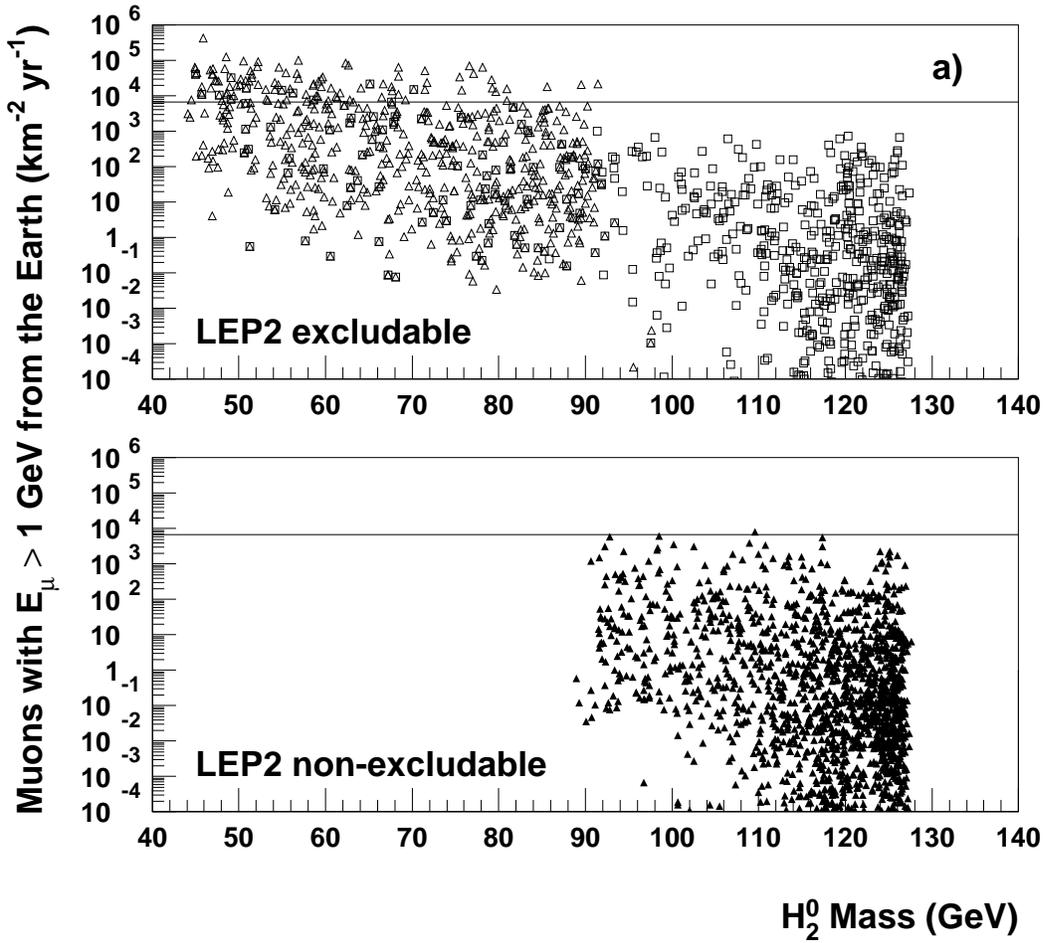,width=\textwidth} 
  \caption{The indirect detection rates from neutralino annihilations
    in a) the Earth and b) the Sun versus $m_{H_2}$\@. The
  horizontal line is the Baksan limit \protect\cite{Baksan}. The upper
  part are models which can be excluded at LEP2 after 150 pb$^{-1}$ of running
  at 192 GeV and the lower part are models which cannot be
  excluded by LEP2\@. The open
  triangles are models that can be excluded due to no Higgs discovery
  and the open squares are models that can be excluded due to no chargino
  discovery. }
  \label{fig:rvsmh2}
\end{figure}

Since the special scan employs fairly low values of the $A$ mass, one
may wonder whether the $A$ or the lightest Higgs boson $H_2$ would be
light enough to be discovered or excluded at LEP2. In particular, one
could expect this for the models that give high neutrino rates, since
a low Higgs mass gives a large spin-independent scattering cross
section of (mixed) neutralinos and therefore large capture and
annihilation rates.  Using the expected exclusion limits for LEP2
given in Ref.~\cite{LEP2} we find, indeed, that the models in our
sample with the highest rates are within reach of LEP2, because of the
large cross section for $e^+e^- \to AH_2$.  Technically this is due to
the fact that these models have a large value of the factor
$\cos^2(\beta-\alpha)$ that governs the $ZAH_2$ coupling (with
$\alpha$ being the CP-even Higgs mixing angle). In
Fig.~\ref{fig:rvsmh2} we show the indirect detection rates versus
$m_{H_2}$: the models that can be probed by LEP2 are shown in the
upper part of the figure, the others in the lower part. 
We have used the combination
of all four LEP experiments and assumed a total integrated luminosity
of 150 pb$^{-1}$ at 192 GeV\@. We have also included the possible LEP2
lower limit of 95 GeV on the mass of the chargino and the bounds
coming from the
$e^+e^- \to ZH_2\to Zb\bar{b}$ process.  LEP2 will probe all models in
our scan with $m_{H_2} \lsim 90$ GeV, and in particular most of the
models already ruled out by Baksan.  On the other hand, there are many
models giving quite large indirect detection rates that cannot be
probed by LEP2, but that would be accessible at a large neutrino
telescope.

In conclusion, indirect dark matter searches and LEP2 probe
complementary regions of the supersymmetric parameter space. Moreover,
direct detection (see \cite{bsg}) is reaching a sensitivity that
allows some models to be excluded, with somewhat different
characteristics than those probed by the other methods.  This
illustrates a nice complementarity between direct detection, indirect
detection and accelerator methods to bound or confirm the minimal
supersymmetric standard model.

%%%%%%%%%%%%%%%%%%%%%%%%%%%%%%%%%%%%%%%%%%%%%%%%%%%%%%%%%%%%%%%%%%%%%
%                        ACKNOWLEDGMENTS
%%%%%%%%%%%%%%%%%%%%%%%%%%%%%%%%%%%%%%%%%%%%%%%%%%%%%%%%%%%%%%%%%%%%%

\section*{Acknowledgments} \label{sec:ack}

This work has been partially supported by the EC Theoretical
Astroparticle Network under contract No.~CHRX-CT93-0120 (Direction
G\'en\'erale 12 COMA). L.B. wants to thank the Swedish Natural Science
Research Council (NFR) for support. P.G. is grateful to the organizers
of the Uppsala Astroparticle Workshop, during which this work was
completed.

%Phys. Lett. B formatting
%\setlength{\baselineskip}{.5\baselineskip}

%%%%%%%%%%%%%%%%%%%%%%%%%%%%%%%%%%%%%%%%%%%%%%%%%%%%%%%%%%%%%%%%%%%%%
%                            REFERENCES
%%%%%%%%%%%%%%%%%%%%%%%%%%%%%%%%%%%%%%%%%%%%%%%%%%%%%%%%%%%%%%%%%%%%%

%%%%%%%%%%%%%%%%%%%%%%%%%%%%%%%%%%%%%%%%%%%%%%%%%%%%%%%%%%%%%%%%%%%%%
%                        FIGURES                              
%%%%%%%%%%%%%%%%%%%%%%%%%%%%%%%%%%%%%%%%%%%%%%%%%%%%%%%%%%%%%%%%%%%%%

\clearpage
\section*{Figures}

\setlength{\parindent}{0cm}

\clearpage
\epsfig{file=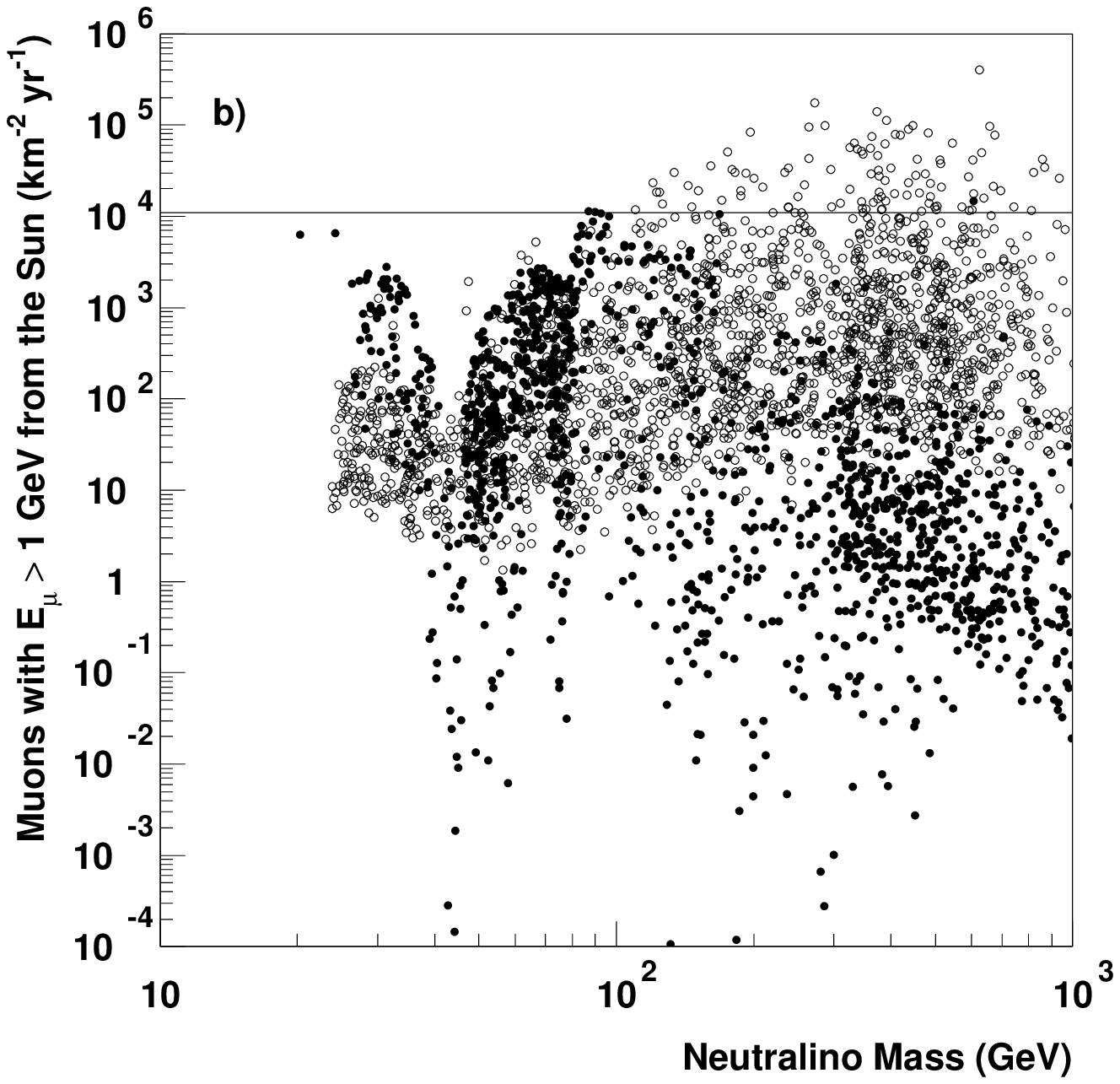,width=\textwidth}

\clearpage

\clearpage
\epsfig{file=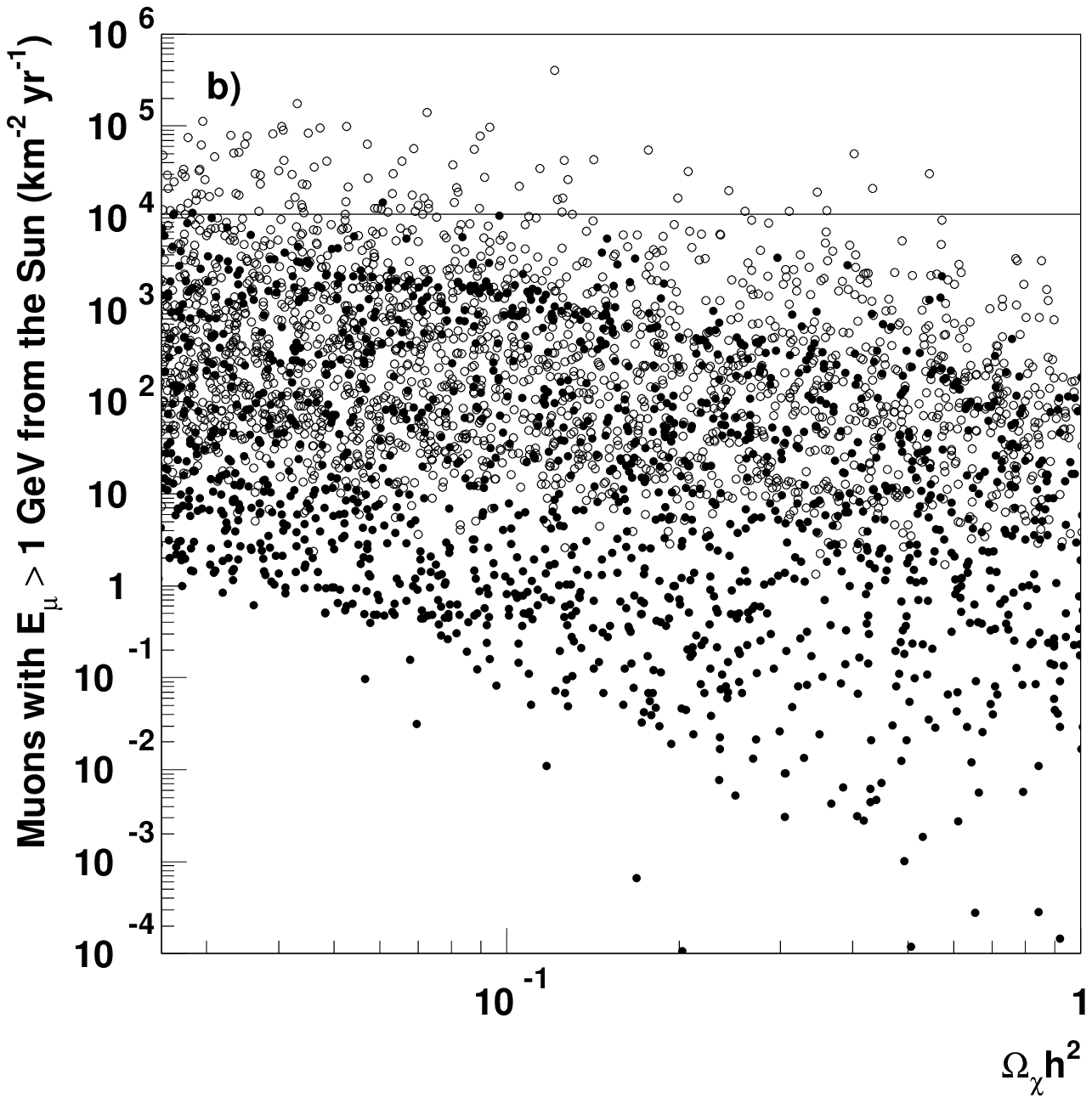,width=\textwidth}

\clearpage

\clearpage

\clearpage
\epsfig{file=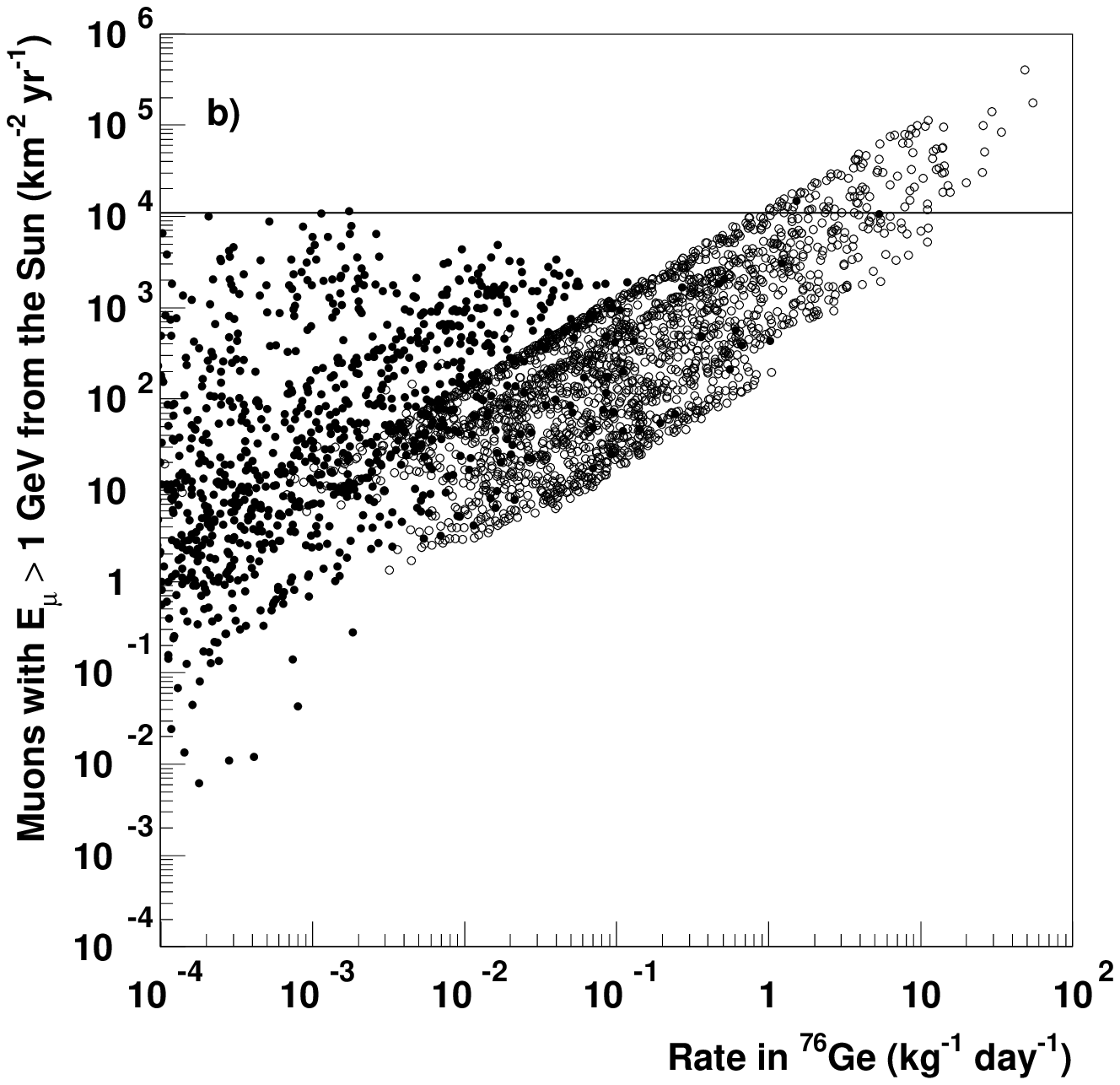,width=\textwidth}

\clearpage

\clearpage
\epsfig{file=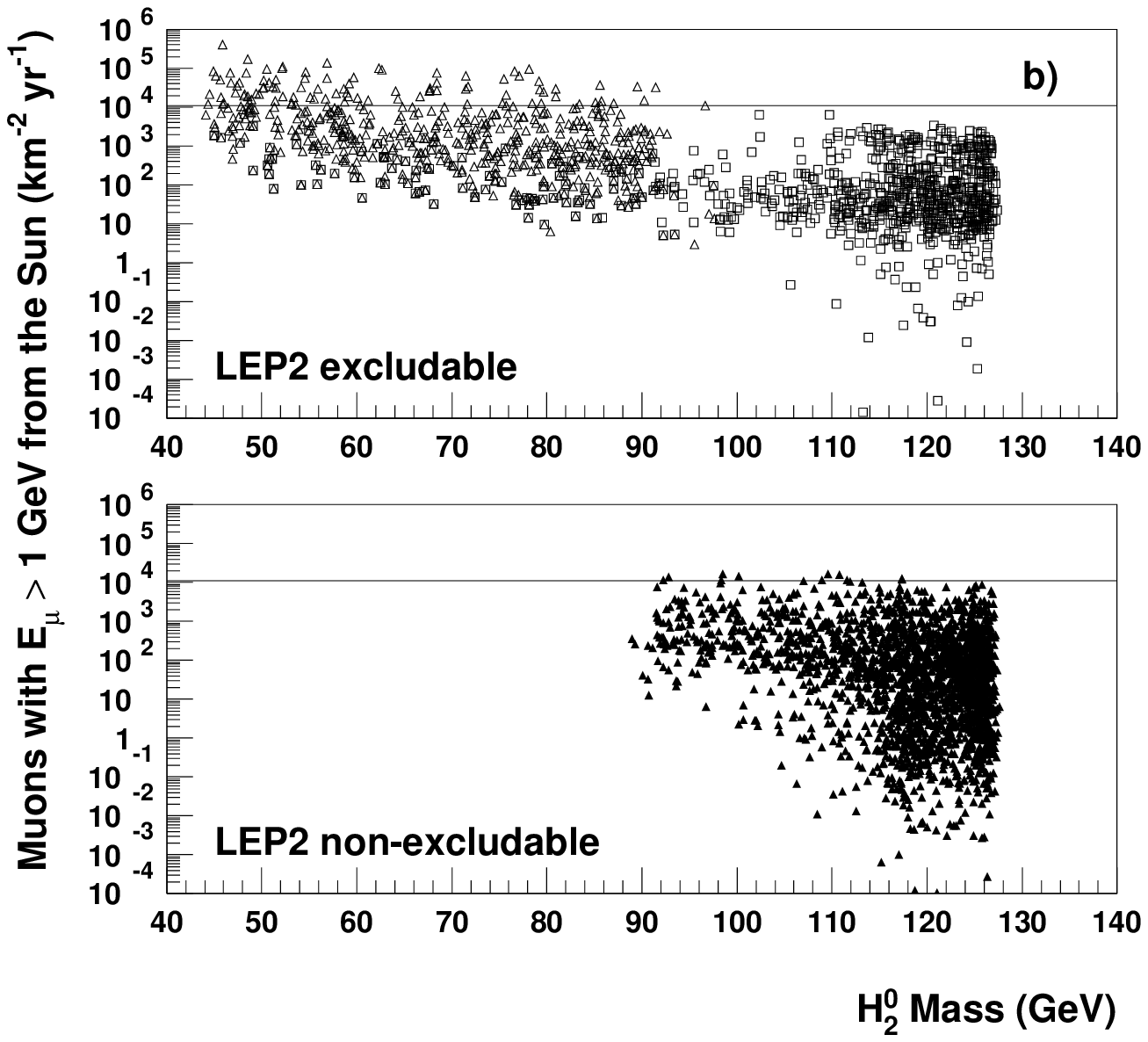,width=\textwidth}

\end{document}